\documentclass[useAMS]{mn2e} 
\usepackage{psfig}

\title[Photometry of binary sdB stars]
{Photometry of four binary subdwarf~B stars and the nature of their unseen
companion stars.}
\author[P. F. L. Maxted et~al.]
       {P. F. L. Maxted$^{1,2}$, T. R. Marsh$^1$, U. Heber$^3$,
L. Morales-Rueda$^1$, R. C. North$^1$ \newauthor
and  W. A. Lawson$^4$ \\ 
$^1$ University of Southampton, Department of Physics \& Astronomy, Highfield, 
 Southampton, S017 1BJ, UK \\
$^2$ Department of Physics, Keele University, Staffordshire, ST5~5BG, UK \\
$^3$ Dr. Remeis-Sternwarte, Astronomisches Institut der Universit\"{a}t
      Erlangen-N\"{u}rnberg, Sternwartstrasse 7, 96049 Bamberg, Germany \\
$^4$ School of Physics, University College UNSW, Australian Defence Force
Academy, Canberra ACT 2600, Australia }
\date{Accepted ---; Received ---}

\pagerange{\pageref{firstpage}--\pageref{lastpage}}
\pubyear{2001}

\newcommand{\Msolar}{\mbox{${\rm M}_{\odot}$}}
\newcommand{\Rsolar}{\mbox{${\rm R}_{\odot}$}}

\newcommand{\vsini}{\mbox{${\rm V}_{\rm rot}\sin i$}}
\newcommand{\logg}{\mbox{$\log g$}}
\newcommand{\kms}{\mbox{${\rm km\,s}^{-1}$}}

\begin{document}

\maketitle

\label{firstpage}

\begin{abstract}
 We present lightcurves of four binary subdwarf~B stars,
Ton~245, Feige~11, PG\,1432+159 and PG\,1017$-$086. We also present new
spectroscopic data for PG\,1017$-$086 from which we derive its orbital period,
$P=0.073$\,d, and the mass function, $f_m= 0.0010\pm0.0002\Msolar$. This is
the shortest period for an sdB binary measured to-date. The values of $P$ and
$f_m$ for the other sdB binaries have been published elsewhere. We are able to
exclude the possibility that the unseen companion stars to Ton~245, Feige~11
and PG\,1432+159 are main-sequence stars or sub-giant stars from the absence
of a sinusoidal signal which would be caused by  the irradiation of such a
companion star, i.e., they show no reflection effect. The unseen companion
stars in these binaries are likely to be white dwarf stars. By contrast, the
reflection effect in PG\,1017$-$086 is clearly seen. The lack of eclipses in
this binary combined with other data suggests that the companion is a low mass
M-dwarf or, perhaps, a brown dwarf. \end{abstract} 

\section{Introduction}
 Subdwarf~B (sdB) stars dominate surveys for extremely blue stars 
brighter than  B$\approx$16 (Green, Schmidt \& Liebert 1986; Downes 1986;
Kilkenny et~al., 1997). Their effective temperatures (T$_{\rm eff}$ =
20\,000K -- 40\,000K) and surface gravities ($\log g$ = 5.5
-- 6.5) place the majority of sdB stars on the extreme horizontal branch
(EHB), i.e., they appear in the same region of the T$_{\rm eff}$\,--\,$\log g$
plane as evolutionary tracks for core helium burning stars with core masses of
about 0.5\Msolar\ and extremely thin ($\la 0.02\Msolar$) hydrogen envelopes
(Heber 1986; Saffer et~al. 1994). The extremely low mass of the hydrogen
envelope in sdB stars is thought to be due to extensive mass loss when the
star was a red giant near the tip of the red giant branch (RGB), i.e., just
prior to ignition of helium in the degenerate helium core. If mass loss occurs
while the red giant is near the tip of the red giant branch, the core can go
on to ignite helium, despite the dramatic mass loss, and may then appear as an
EHB star (d'Cruz et~al. 1996). The cause of the extensive mass loss has been a
matter of some debate, but a recent survey for short period binary EHB stars
by Maxted et~al. (2001) has shown that in at least 2 out of 3 EHB stars,
interactions with companion stars are responsible. In this scenario, the
expanding red giant star comes into contact with its Roche lobe and begins to
transfer mass to its companion star. This mass transfer is highly unstable, so
a ``common-envelope'' forms around the companion and the core of the red
giant. The drag on the companion orbiting inside the common envelope  leads to
extensive mass loss and dramatic shrinkage of the orbit (Iben \& Livio 1993).

 The properties of sdB stars, e.g., their orbital period distribution, are a
strong test of population synthesis models for binary stars because the common
envelope phase which produced the sdB star must have occurred in a star with a
degenerate helium core at the tip of the red giant branch. This places useful
limits on the mass and radius of the star at the onset of the common envelope
phase. Another property of these binaries which can, in principle, be compared
to population synthesis models is the relative number of degenerate and
non-degenerate companions, i.e., the fraction of sdB stars with main-sequence
or sub-giant companions compared to the fraction with white dwarf companions.
In this paper we outline a simple method to determine the nature of the
companion in practice and apply the method to four sdB binaries with known
orbital periods.

\section{The method}
 In those cases where the companion to an sdB star is not seen directly in the
optical spectrum, some other method must be found to determine the nature of
the companion star. A sensitive method to detect main-sequence or sub-giant
companions is to obtain an accurate lightcurve and to look for the effect of
the irradiation of one side of the companion star by the hot sdB star. This
produces  an easily detectable signal with the same period as the orbital
period with an approximately sinusoidal shape because the binary appears
brightest when we see the irradiated hemisphere face-on and is fainter half an
orbit later when we see more of the non-irradiated hemisphere. This sinusoidal
distortion to the lightcurve is known as the reflection effect. 

\subsection{The reflection effect.}
It is useful to estimate the amplitude of the reflection effect we expect from
a cool, faint companion to an sdB star using the following assumptions and
approximations. If the sdB star has an effective temperature $T_1$ and a
radius $R_1$ the flux intercepted by a companion star of radius $R_2$ at a
distance $a$, is $\pi R_2^2 \sigma T_1^4 (R_1/a)^2$, where $\sigma$ is the
Stefan-Boltzmann constant. If all of the irradiating flux is re-emitted by the
heated face, the effective temperature of the heated face is $T_h\approx T_1
(R_1/\sqrt{2}a)^{\frac{1}{2}}$, where we have assumed the intrinsic luminosity
of the companion star is negligible. The amplitude of the reflection effect at
a given wavelength depends on the spectrum of the sdB star and the spectrum of
the reprocessed light from the heated face of the companion star. The spectrum
of the reprocessed light we observe will change with orbital phase because the
temperature varies across the heated face and the light will not be radiated
isotropically from the heated face. The model we describe later takes account
of these effects, but for the purposes of estimating the amplitude of the
reflection effect we can calculate the maximum flux observed from the heated
star using the approximation that it appears as a star of radius $R_2$ and
effective temperature of $T_h$.  For observations at most optical and
near-infrared wavelengths, the Rayleigh-Jeans limit of a black body spectrum is
a good approximation for the spectrum of the sdB star, i.e, the intensity is
proportional to the effective temperature. In the case of strong irradiation,
$T_h$ will also be sufficiently high for this approximation to apply to the
spectrum of the heated face of the companion.

 If we consider a binary which is seen nearly edge-on but is not eclipsing, we
find that the difference in magnitudes when we see the irradiated and
non-irradiated faces $\delta m$, is 
\begin{eqnarray*}
 \delta m & \approx & 2.5 \log_{10} \left[ 1 + \left(\frac{R_2}{R_1}\right)^2
\left(\frac{R_1}{\sqrt{2}a}\right)^{\frac{1}{2}} \right]  \\
          & \approx & \left(\frac{R_2}{R_1}\right)^2
\left(\frac{R_1}{a}\right)^{\frac{1}{2}} {\rm ~~if~\delta m~is~small}
\end{eqnarray*}

 Although this expression is approximate, it does enable us to estimate the
amplitude of the reflection effect we might expect from an sdB star with a
main-sequence companion. More importantly, it shows that $\delta m \propto
R_2^2$, so a typical white dwarf with a  radius of 0.01\Rsolar\ will produce a
reflection effect at least 100 times less than a typical M-dwarf with a radius
of 0.1\Rsolar. The actual difference will be much larger than this if the
white dwarf is not exceptionally cool but has a more typical  effective
temperature for white dwarfs of 10--20\,000K because the temperature contrast
between the heated and unheated hemispheres would then be much less. For
example, an sdB star with $T_{\rm eff} = 30\,000$K and a radius of 0.2\Rsolar\
with a cool companion 1\Rsolar\ distant (P$\sim$ 4\,h) will show a reflection
effect of about 0.1\,magnitudes, which is easily detectable with differential
CCD photometry. By contrast, a white dwarf with a radius of 0.01\Rsolar\ at
the same distance gives rise to a reflection effect of no more than
0.001\,magnitudes. Even if the inclination of the binary reduced the amplitude
of the reflection effect by an order of magnitude, it would be straightforward
to distinguish the nature of the companions in these two hypothetical binary
sdB stars using the presence or absence of a reflection effect in the
lightcurve.

\subsection{Our model.}
 We have used a simple model to produce synthetic lightcurves for the sdB
binaries we have observed which is based on numerical integration over the
visible surface of the heated star. The visible surface of the irradiated star
is defined by a Roche potential. The effective temperature at the integration
points on the heated face, $T_h^{\prime}$ is calculated from $T_h^{\prime} =
(T_u^4 + f^{\prime}/\sigma)^{1/4}$, where $T_u$ is the mean effective
temperature of the unheated face and $f^{\prime}$ is the irradiating flux over
that region of the star allowing for the angle between the normal to the
surface and the direction of the sdB star. The other parameters of our model
are the radius of the sdB star in units of the orbital separation, $r_1$; the
radius of the companion star in units of the distance between its centre and
the inner Lagrangian point measured along the same axis (the filling-factor),
$f$; the inclination of the orbit, $i$; the mass ratio, $q=M_2/M_1$;  the
orbital period, $P$, the effective temperature of the sdB star, $T_1$.
The observed flux at a given wavelength and orbital phase can then be
estimated by assuming that both stars radiate as black-bodies.  We also
include limb darkening and gravity darkening in the calculation of the
reflection effect. The details of the treatment of these effects in our model
has a negligible effect on the predicted amplitude of the reflection effect so
we do not describe them here. Our model does not include any contribution to
the lightcurve due to the distortion of the sdB star by its companion, i.e,
the ellipsoidal effect from the sdB star is ignored. This effect  is
negligible for Ton~245, Feige~11 and PG\,1432+159 but not for PG\,1017$-$086.

 Our calculation of the effective temperature over the surface of the
companion star is equivalent to assuming that the bolometric albedo of the
companion star, $\alpha$,  is 1. There is good observational evidence for cool
companions to hot subdwarf stars having high bolometric albedos, at least for
short period binaries. For example, Drechsel et~al. (2001) have presented B
and R lightcurves of the eclipsing sdB star HS\,0705+6700 which has an M-dwarf
companion and an orbital period of $P=2.3h$. They used the Wilson-Devinney
code (Wilson \& Devinney 1971) to model the lightcurve and found that to
achieve satisfactory fits to the lightcurves a value of $\alpha=1$ is
required. The fits were improved by using a value of $\alpha=1.1$, which is
unrealistic and is a consequence of using a monochromatic lightcurve model
based on black body radiation to model a broad band lightcurves of an
irradiated star whose spectrum is certainly very different to a black body.
Our model is also affected by this problem. Similarly, Hilditch, Harries \&
Hill (1996) analysed the lightcurves of HW~Vir, KV~Vel (sdO+M, P=$8.6h$) and
AA Dor (sdO+M, $P=6.3h$) with the lightcurve model {\sc light2} (Hill \&
Rucinski 1993) and also found $\alpha=1$ or more is required to reproduce the
reflection effect. Wood, Zhang \& Robinson (1993) confirmed the requirement
for a high albedo in HW~Vir by using the Wilson-Devinney code to analyse their
UBVR lightcurves of HW~Vir.

 Although our model is quite simple, it is able to predict the amplitude of
the reflection effect in sdB binaries with faint companions with an accuracy
of about 50 percent. To demonstrate this, we have calculated the amplitude of
the reflection effect for two sdB stars with faint companions, HW~Vir and
PG\,1336$-$018 and the sdOB binary V477~Lyr. These three stars are eclipsing
binaries so the properties of the stars can be determined independently. This
is shown in Table~\ref{OtherTable}, where we  compare
the observed amplitude of the reflection effect in the V and I bands,
$\delta$V and $\delta$I, to the value calculated using our model. We also list 
the properties of each binary in Table~\ref{OtherTable} 
which have the greatest effect on the values of $\delta$V and $\delta$I
predicted by our model. In all three cases we see that the amplitude in the V
band is predicted correctly to within 50 percent. For HW~Vir where the
amplitude has also been measured in the I-band, the amplitude predicted by our
model agrees very well with the observed value. 

\begin{table*}
\caption{\label{OtherTable}
The observed amplitude of the reflection effect in HW~Vir, PG\,1336-018 and
V477~Lyr and the amplitude calculated from our model.}
\begin{tabular}{@{}lrlrrrrrrrrrrrr}
&&&&&&&&&\multicolumn{2}{c}{$\delta$V} &\multicolumn{2}{c}{$\delta$ I} \\
Name&\multicolumn{1}{l}{P(d)} &\multicolumn{1}{l}{Spec. Type}& 
\multicolumn{1}{c}{T$_1$(K)}& \multicolumn{1}{l}{M$_1$} &
\multicolumn{1}{c}{q}&\multicolumn{1}{l}{$i$($^{\circ}$)}& 
\multicolumn{1}{c}{f}& \multicolumn{1}{l}{$r_1$}&
\multicolumn{1}{l}{Obs.}&\multicolumn{1}{l}{Cal.}&
\multicolumn{1}{l}{Obs.}&\multicolumn{1}{l}{Cal.}& Ref. \\
\hline
HW Vir     &0.117& sdB+M  & 28500 & 0.5 &0.3  & 81 &0.34& 0.205 &0.26 & 0.24&0.30 &0.30& 1,2 \\
PG\,1336-018 &0.101& sdB+M5 & 33000 & 0.5 &0.3 & 81 &0.53 &0.19 & 0.20 &0.27&--- & --- & 3\\
V477 Lyr & 0.471 & sdOB+M & 60000 & 0.51 & 0.29 & 80.5 & 0.54 & 0.077 & 0.78&
0.65&---&---&4 \\
\hline
\multicolumn{13}{l}{ 1.~Wood \& Saffer (1999); 2.~Kiss et~al. (2000);
3.~Kilkenny et~al. (1998); 4.~Pollacco \& Bell (1994); }\\
\end{tabular} \\ 

\end{table*}

\section{Observations and reductions}

\subsection{Photometry}
  We used the 1m Jacobus Kapteyn telescope (JKT) on the Island of La Palma to
obtain V images and I band images of PG\,1432+159 (57 V images, 52 I
images), Feige~11 (80 V images, 85 I images) and Ton~245 (29 V images, 40 I
images) on the nights 1998 Aug 2--10. Additional observations of Ton~245 (9 V
images, 9 I images) were acquired on the night 1999 Jan 3. The detector used
was a  TEK charged-coupled device (CCD) with $1024^2$ pixels giving an image
scale of 0.34\arcsec\ per pixel. Exposure times varied between 10s and 120s
and the deadtime between exposures was 60s. We obtained 40 V band images of
PG\,1017$-$086 with the same telescope using a SITe CCD with $2048^2$ pixels
giving an image scale of 0.34\arcsec\ per pixel on the night 2001 May 3. The
exposure times were 70s or 100s and the  deadtime between exposures was 110s.
We also acquired 45 V band images of PG\,1017$-$086 with the SAAO 1m telescope
using a SITe CCD with $1024^2$ pixels giving an image scale of 0.31\arcsec\
per pixel on the night 2001 May 7. The exposure time was 90s and the deadtime
between exposures was 73s.

\begin{table}
\caption{\label{PhotTable} Position and magnitudes of the target and 
comparison stars used for our differential photometry.}
\begin{tabular}{lrrrr}
Target  &
\multicolumn{1}{c}{$\alpha$(J2000)}&\multicolumn{1}{c}{$\delta$(J2000)} & 
\multicolumn{1}{c}{V} & Ref.\\
\hline
Ton~245        & 15 40 35.4 &+26 47 42  & 13.89  & 1 \\
(N1330313366)  & 15 40 43.1 &+26 46 52  & 15.79 \\
(N133031310095)& 15 40 37.4 &+26 48 21  & 18.17 \\
(N133031310109)& 15 40 32.5 &+26 48 43  & 17.36 \\
\\                         
Feige~11       & 01 04 21.7 &+04 13 37  & 12.06 & 2 \\
(N320132243)   & 01 04 28.0 &+04 11 55  & 14.47 \\
(N320132247)   & 01 04 28.4 &+04 11 25  & 13.80 \\
\\                         
PG\,1432+159   & 14 35 19.2 &+15 40 14 & 13.90 & 3\\
(N1313121947)  & 14 35 16.0 &+15 39 59 & 15.94 \\
(N1313121769)  & 14 35 13.7 &+15 36 41 & 15.61 \\
(N1313121359)  & 14 35 11.9 &+15 36 45 & 14.82 \\
(N13131213083) & 14 35 26.3 &+15 39 17 & 16.55 \\
\\                         
PG\,1017$-$086 & 10 20 14.5 &$-$08 53 46 & 14.43 & 3\\
(S1212232199)  & 10 20 08.5 &$-$08 50 58 & 13.11 \\
\hline
\multicolumn{5}{l}{1. Iriarte (1959); 2. Landolt (1983); 3. Wesemael et~al.
(1992).}\\
\end{tabular}
\end{table}

 The bias level in every image was determined from the overscan regions and
was subtracted from the image before further processing. Images of the
twilight sky devoid of any bright stars were used to determine flat-field
corrections by forming the median image of 3--5 twilight sky images in each
filter, one for each night's data. We used optimal photometry (Naylor 1998) to
determine instrumental magnitudes of the stars in each frame. We checked for
variability in the stars other than the target star in each frame before
calculating differential magnitudes between the target star and the total flux
in the comparison stars. The positions and approximate V magnitudes of the
targets and comparison stars are given in Table~\ref{PhotTable}. The
coordinates and identification numbers (in parentheses)
 in Table~\ref{PhotTable} were taken from Guide
Star Catalogue-II 
 \footnote{
   The Guide Star Catalogue-II is a joint project of the Space Telescope
    Science Institute and the Osservatorio Astronomico di Torino. Space
    Telescope Science Institute is operated by the Association of
    Universities for Research in Astronomy, for the National Aeronautics
    and Space Administration under contract NAS5-26555. The participation
    of the Osservatorio Astronomico di Torino is supported by the Italian
    Council for Research in Astronomy. Additional support is provided by
    European Southern Observatory, Space Telescope European Coordinating
    Facility, the International GEMINI project and the European Space
    Agency Astrophysics Division.}.
 We have used published Str\"{o}mgren
$y$ magnitudes for PG\,1432+159 and PG\,1017$-$086 in place of V, the
difference for these blue stars are small ($\loa 0.05$mag). The magnitude of
the comparison stars given in  Table~\ref{PhotTable} was calculated from the
published target magnitude given and the median magnitude difference from our
own photometry.

\subsection{Spectroscopy}

 We obtained low-resolution spectra of Ton~245 using  the red arm of the ISIS
double beam spectrograph on the 4.2m William Herschel Telescope on the Island
of La Palma on the night 2001 February 22. We also obtained spectra with the
same instrument of the K3V star GL\,250\,A and the M1.5V star GL\,220. We used a
158 line/mm grating and a TEK CCD with a 1\,arcsec wide slit to obtain spectra
with a resolution of  5--6\AA\  and a mean dispersion of 2.9\AA/pixel. We
applied a flux calibration to these spectra using observations of G191$-$B2B
and the tabulated fluxes of Oke (1990). We have made no correction for
slit-losses in this calibration.

 We observed PG\,1017$-$086 using with the 2.5m Isaac Newton Telescope on the
Island of La Palma. A total of 22 spectra were acquired on the nights 2000
April 11, 2001 March 8--11 and 2001 May 6. Spectra were obtained with the
intermediate dispersion spectrograph  using the 500mm camera, a 1200~line/mm
grating and a TEK CCD as a detector. The spectra cover 400\AA\ around the
H${\alpha}$ line at a dispersion of 0.39\AA\ per pixel. The slit width used
was 0.97\,arcsec which gave a resolution of about 0.9\AA. The exposure time
per spectrum was 600s or 900s. We also obtained a continuous series of 26
spectra of PG\,1017$-$086 on the night 2001 March 12 using the 235mm camera, a
900~line/mm grating and an EEV CCD. The spectra cover the wavelength range
3700\,--\,5400\AA\ at a dispersion of 0.63\AA\ per pixel with a resolution of
about 1.6\AA\ and the exposure time per spectrum was 300s. 

 We extracted the spectra from the images using optimal extraction to maximize
the signal-to-noise (Marsh 1989). We observed arcs before and after all
observations of PG\,1017$-$086. The arcs associated with each stellar
spectrum were extracted using the same weighting determined for the stellar
image to avoid possible systematic errors due to the tilt of the spectra on
the detector. The wavelength scale was determined from a fit to measured arc
line positions and in each case the standard deviation of the fit is much less
than 1 pixel. The wavelength scale for an individual spectrum was determined
by interpolation to the time of mid-exposure from the fits to arcs taken
before and after the spectrum to account for the small amount of drift in the
wavelength scale due to flexure of the instrument. Statistical errors on every
data point calculated from photon statistics are rigorously propagated through
every stage of the data reduction.
 
\section{PG\,1017$-$086}

\subsection{Radial velocities}
 We first observed PG\,1017$-$086 as part of a survey for binary sdB stars
(Maxted et~al. 2001). The two spectra we obtained for that survey showed a
change in radial velocity measured from the H$\alpha$ line of about 60\kms\ 
over half an hour.  We obtained further observations of the H$\alpha$ line of
PG\,1017$-$086 to determine the orbital period and mass function. To measure
the radial velocity we used least-squares fitting of a model line profile.
This model line profile is the summation of three Gaussian profiles with
different widths and depths but with a common central position which varies
between spectra.  Only data within 2000\,km\,s$^{-1}$ of the H${\alpha}$ line
is included in the fitting process and the spectra are normalized  using a
linear fit to the continuum either side of the H${\alpha}$ line. We used a
least-squares fit to one of the spectra to determine an initial shape of the
model line profile. A least squares fit of this profile to each spectrum in
which the position of the line is the only free parameter gives an initial set
of radial velocities. We used these initial radial velocities to fix the
position of the H${\alpha}$ line in a simultaneous fit to all the spectra to
obtain an improved model line profile. A least squares fit of this profile to
each spectrum yields the radial velocities given in Table~\ref{RVTable} and
are labeled ``Red''. A similar process was used to measure radial velocities
from the H$\beta$ -- H$\epsilon$ lines with two Gaussian profiles used to
model each line. These are given in Table~\ref{RVTable} and are labeled
``Blue''.  The uncertainties quoted are calculated by propagating the
uncertainties on every data point in the spectra right through the data
reduction and analysis. 

\begin{figure} 
\caption{\label{RVFitFig} Radial velocities of PG\,1017$-$086 measured from
the H$\alpha$ line (open circles) and from the H$\beta$ -- H$\epsilon$
 lines (closed circles). The sine wave fit described in the text is also shown
(solid line). }
\psfig{file=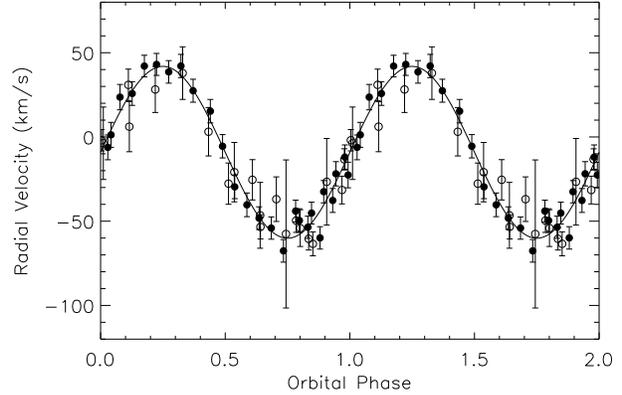,width=0.45\textwidth} 
\end{figure}

 The periodogram of the radial velocities given in Table~\ref{RVTable} shows a
single unambiguous orbital frequency at 13.70 cycles/day. We used a
least-squares  fit of a sine wave of the form $\gamma + K\sin((T-T_0)/P)$ to
obtain the parameters given in Table~\ref{RVFitTable}, where $\gamma$ is
the systemic velocity, $K$ is the projected orbital speed, $T$ is the time of
mid-exposure of the spectrum and $P$ is the orbital period. The measured
radial velocities are shown in Fig.~\ref{RVFitFig} as a function of orbital
phase together with the sine wave determined from the least-squares fit.
 
\begin{figure} 
\caption{\label{BlueFitFig} Synthetic spectrum fit to the blue Balmer lines
and helium lines of PG\,1017$-$086}
\psfig{file=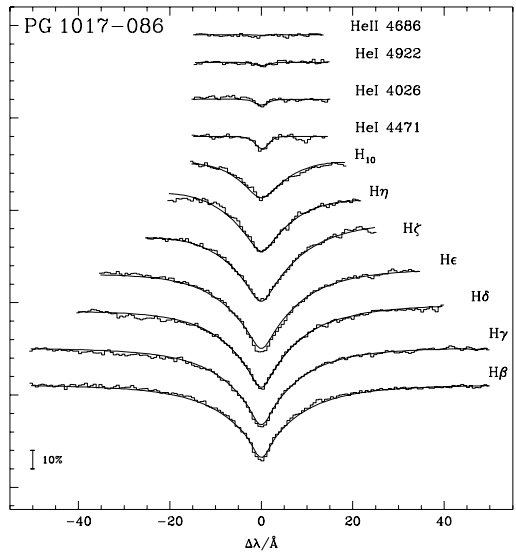,width=0.45\textwidth}
\end{figure}

 We rebinned all the spectra near H$\alpha$ onto a common wavelength scale
allowing for the measured radial velocity shifts and then formed the average
spectrum, and similarly for the spectra of the bluer Balmer lines. There are
no spectral features attributable to a cool companion star visible in either
of these average spectra. We used the average blue spectrum to measure the
effective temperature, T$_{\rm eff}$, the surface gravity $\log g$ and the
helium abundance by number, $y$, by fitting model spectra to the Balmer lines
(H$\beta$ to H$\,$10), the He~I lines (4026\AA, 4388\AA, 4471\AA, 4713\AA,
4922\AA) and He~II 4686\AA\ lines using the procedure outlined in Saffer
et~al. (1994). We used the synthetic spectra derived from H and He line
blanketed NLTE model atmospheres of Napiwotzki (1997).  We find T$_{\rm eff} =
30300 \pm 80$K, $\logg =  5.61 \pm 0.02$ and $y=0.0016 \pm 0.0001$ from these
fits, where the uncertainties are ``internal errors'' from the fitting
procedure and do not include uncertainties in the models themselves. The
spectrum and fit are shown in Fig.~\ref{BlueFitFig}. The synthetic spectra
were convolved beforehand with a Gaussian profile of the appropriate width to
account for the instrumental profile and with a broadening function to account
for a projected rotational velocity, \vsini, of 118\kms. This was determined
from a least-squares fit to the H$\alpha$ line of a synthetic line profile for
the appropriate T$_{\rm eff}$, \logg\  and $y$ values convolved with a
broadening function for various values of \vsini. From a plot of $\chi^2$
versus \vsini\ we estimate a value of \vsini=$118\stackrel{+7}{\scriptstyle
-4}$\,\kms. The radius of a 0.5\Msolar\, sdB star with \logg=5.61 is
0.19\Msolar. If we assume tidal forces have forced the sdB star to co-rotate
with the binary, the rotational velocity of the sdB star is 132\kms. The
measured value of \vsini\ would then imply that the inclination of the binary
is $63^{\circ}\stackrel{+8^{\circ}}{\scriptstyle -4^{\circ}}$. 

\begin{table}
\caption{\label{RVTable} Measured heliocentric radial velocities for
PG\,1017$-$086.}
\begin{tabular}{@{}rrr}
\multicolumn{1}{l}{HJD$-$2450000} & \multicolumn{1}{l}{Radial velocity
(km\,s$^{-1}$)} & Spectrum \\ 
 1646.4502 & $ -63.4 \pm  7.1$ & Red  \\
 1646.4614 & $  -1.9 \pm  6.4$ & Red  \\
 1977.4525 & $ -27.7 \pm 12.3$ & Red  \\
 1977.4595 & $ -25.4 \pm 12.0$ & Red  \\
 1977.4665 & $ -36.9 \pm 13.2$ & Red  \\
 1977.4735 & $ -54.1 \pm 10.9$ & Red  \\
 1978.5806 & $ -31.5 \pm  8.2$ & Red  \\
 1978.5910 & $  30.9 \pm  9.3$ & Red  \\
 1979.5198 & $ -60.5 \pm  6.6$ & Red  \\
 1979.5302 & $ -13.3 \pm  6.3$ & Red  \\
 1979.5786 & $ -53.4 \pm  7.2$ & Red  \\
 1979.5891 & $ -49.6 \pm  7.0$ & Red  \\
 1980.4757 & $ -37.7 \pm  7.1$ & Blue \\
 1980.4793 & $ -12.0 \pm  7.2$ & Blue \\
 1980.4829 & $  -6.2 \pm  7.4$ & Blue \\
 1980.4864 & $  23.6 \pm  7.5$ & Blue \\
 1980.4900 & $  25.8 \pm  7.0$ & Blue \\
 1980.4936 & $  42.1 \pm  6.4$ & Blue \\
 1980.4971 & $  43.1 \pm  6.5$ & Blue \\
 1980.5007 & $  38.6 \pm  6.7$ & Blue \\
 1980.5043 & $  42.1 \pm  6.9$ & Blue \\
 1980.5078 & $  27.4 \pm  6.8$ & Blue \\
 1980.5129 & $  15.3 \pm  7.0$ & Blue \\
 1980.5164 & $  -5.5 \pm  6.9$ & Blue \\
 1980.5200 & $ -29.6 \pm  7.0$ & Blue \\
 1980.5236 & $ -40.3 \pm  7.0$ & Blue \\
 1980.5272 & $ -48.2 \pm  6.7$ & Blue \\
 1980.5307 & $ -54.1 \pm  6.6$ & Blue \\
 1980.5343 & $ -67.6 \pm  6.7$ & Blue \\
 1980.5379 & $ -44.0 \pm  6.5$ & Blue \\
 1980.5414 & $ -53.6 \pm  6.6$ & Blue \\
 1980.5450 & $ -59.9 \pm  6.7$ & Blue \\
 1980.6119 & $ -49.6 \pm  7.0$ & Blue \\
 1980.6155 & $ -45.2 \pm  6.5$ & Blue \\
 1980.6191 & $ -32.5 \pm  6.6$ & Blue \\
 1980.6226 & $ -21.9 \pm  7.2$ & Blue \\
 1980.6262 & $ -22.6 \pm  7.6$ & Blue \\
 1980.6298 & $   1.3 \pm  7.4$ & Blue \\
 2036.3872 & $ -26.6 \pm 25.7$ & Red  \\
 2036.3948 & $  -3.7 \pm 21.5$ & Red  \\
 2036.4024 & $   6.1 \pm 14.9$ & Red  \\
 2036.4100 & $  28.2 \pm 13.6$ & Red  \\
 2036.4180 & $  37.9 \pm 15.6$ & Red  \\
 2036.4256 & $   3.1 \pm 14.4$ & Red  \\
 2036.4332 & $ -20.9 \pm 17.6$ & Red  \\
 2036.4407 & $ -46.5 \pm 19.6$ & Red  \\
 2036.4483 & $ -57.6 \pm 44.0$ & Red  \\
\end{tabular}
\end{table}

\subsection{The lightcurve and the nature of the companion}

\begin{figure} 
\caption{\label{LC1017Fig} The V-band lightcurve of the PG\,1017$-$086 observed
with the SAAO 1m telescope (filled circles) and with the JKT (open circles).
The solid lines are cosine functions with an amplitude of 0.083\,magnitudes}
\psfig{file=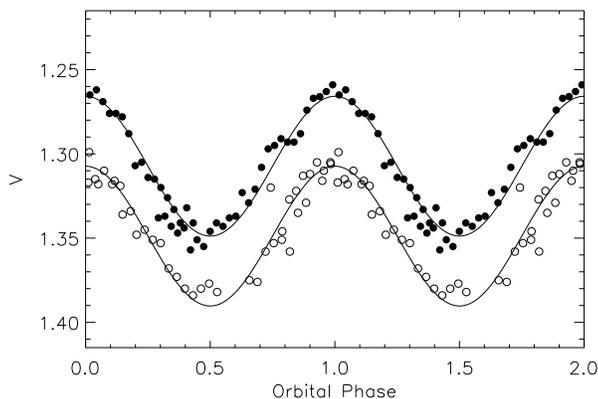,width=0.45\textwidth}
\end{figure}

 Our photometry of PG\,1017$-$086 is shown in Fig.~\ref{LC1017Fig} as a
function of the orbital phase calculated from the values of $T_0$ and $P$
derived above. The reflection effect with an amplitude of about
0.08\,magnitudes can be clearly seen and  maximum light is at phase zero as
expected, i.e., when the sdB star is closest to the observer. The difference
between the mean level of the lightcurves observed with the SAAO 1m telescope
and the JKT is due to a difference in the sensitivity of the CCDs used
at the bluest wavelengths passed by the V filter. This makes PG\,1017$-$086
appear brighter than its redder comparison star for the CCD with better blue
sensitivity.

 The minimum mass of the companion to a 0.5\Msolar\, sdB star for the values of
$P$ and $K$ observed in PG\,1017$-$086 is 0.0687$\pm$0.025\Msolar. The mass of
the companion for the inclination of
$63^{\circ}\stackrel{+8^{\circ}}{\scriptstyle -4^{\circ}}$ calculated above is
$0.078\stackrel{+0.005}{\scriptstyle -0.006}\Msolar$.  If we assume the
companion is a low mass M-dwarf with a typical radius of 0.085\Rsolar, we find
that our simple model for the reflection effect predicts an amplitude for the
lightcurve of 0.074\,magnitudes, which is consistent with the amplitude
observed (Fig.~\ref{LC1017Fig}) given the uncertainties involved, i.e., about
50\,percent. The amplitude of the reflection effect seen in the lightcurve
cannot be produced by a white dwarf companion. The lack of any eclipse sets an
upper limit to the inclination of 72$^{\circ}$ for the radius of sdB star we
have calculated. The limit reduces to 63$^{\circ}$ if we assume the companion
has a radius of 0.085\Rsolar, which is consistent with the inclination used in
the calculation of the reflection effect. Our model does not include the
ellipsoidal effect due to the sdB star, which is expected to be about
0.01\,magnitudes. This is too small to be measured reliably from our
lightcurves, but would be measurable with improved data, particularly at bluer
wavelengths where the sdB star dominates. The size of the ellipsoidal effect
depends strongly on the size of the star relative to its Roche lobe, so this
measurement would usefully constrain the properties of the stars in
PG\,1017$-$086.

 In summary, we can say the companion to PG\,1017$-$086 is a low mass
star or, perhaps, a brown dwarf, but is certainly not a white dwarf star.

\begin{table}
\caption{Circular orbit fit to the measured radial
velocities for PG\,1017$-$086.\label{RVFitTable} }
\begin{tabular}{@{}lr@{}}
$T_0$ (HJD)&   2452036.3940  $\pm$ 0.0005 \\
$P$ (days) &      0.0729938  $\pm$ 0000003 \\
$ \gamma(\kms)$ & $-$9.1 $\pm$ 1.3 \\
$K(\kms)$  &  51.0 $\pm$ 1.7 \\
$\chi^2$ & 46.1 \\
N &  47 \\
\end{tabular}
\end{table}

\section{Ton~245, Feige~11 and  PG\,1432+159}

In this section we have to go through a fairly complex chain of logic in order
to establish the final result, which is that the companions of the three stars
Ton~245, Feige~11 and PG\,1432+159 must all be compact as opposed to
main-sequence stars or brown dwarfs. To help the reader, we now give an
overview of the reasoning we employ.

We detect no significant reflection effect in any of these stars and our task
is to show that this is enough to say that they must have compact companions,
most likely white dwarfs. The complicating factor is the unknown orbital
inclination which also affects the reflection effect amplitude in the sense
that the amplitude becomes small as orbits become more face-on. In fact the
reflection amplitude can also decrease at {\em high} inclinations because for
a given radial velocity amplitude a high inclination implies a reduced
companion star mass which, in turn, implies a reduced radius and, therefore, a
reduced reflection effect. However it is the reduction at low inclinations
that is much more significant, e.g., see Fig.~\ref{AmpFig}. This essentially
means that very low mass companions are ruled out by the radial velocity
amplitudes and therefore we cannot get a small reflection effect by appealing
to brown dwarf companions for these three stars. Thus, the problem boils down
to ruling out low inclinations as a way of getting low amplitudes. We can do
so as follows:

From spectra of the targets we can place upper limits upon the contribution of
the companion stars to the sdB spectra. Given the luminosity of the sdB stars
at that wavelength,  the main-sequence mass-luminosity relation implies upper
limits upon the companion star masses. For a given sdB mass, radial velocity
amplitude and orbital period, an {\em upper} limit on the companion mass gives
a {\em lower} limit upon the inclination (see Fig.~\ref{AmpFig}). This in turn
gives a lower limit upon the predicted amplitude of the reflection effect on
the assumption of main-sequence companions. The loop is finally closed when we
establish that this lower limit should have been detectable in each case, and
therefore that the companions must be compact.

\subsection{Upper limits to the luminosity ratios.}

 In order to set upper limits on the contribution from any cool companion star
to the spectra of Ton~245, Feige~11 and  PG\,1432+159, we have considered a
region of the spectrum where we expect to see no features from the sdB star or
the Earth's atmosphere. We then subtract off varying amounts of cool star
spectrum to find the luminosity ratio at which the combined spectrum no longer
looks featureless, i.e., for which a low-order polynomial is no longer a good
fit.

\begin{figure*} 
\caption{\label{LratioFig} The upper limits to the luminosity ratio, $L$, for
Ton~245, Feige~11 and  PG\,1432+159. The panels on the left show the
$\chi^2$ statistic for a least-squares fit of a parabola to the residuals
after subtracting a fraction $L$ of a cool star spectrum from the observed
spectrum of the sdB star. The value of the $\chi^2$ statistic for $L=0$,
$\chi^2_0$ is show by a horizontal solid line and the value $\chi^2_0$+3 which
sets the upper limit to $L$ is shown with a horizontal dashed lines. The
panels on the right show spectra of Feige~11 and PG\,1432+159 together with
the spectrum of the K5V star GL\,69 offset by $-$0.2 for clarity and similarly
for Ton~245 and the K3V star GL\,250\,A.}
\psfig{file=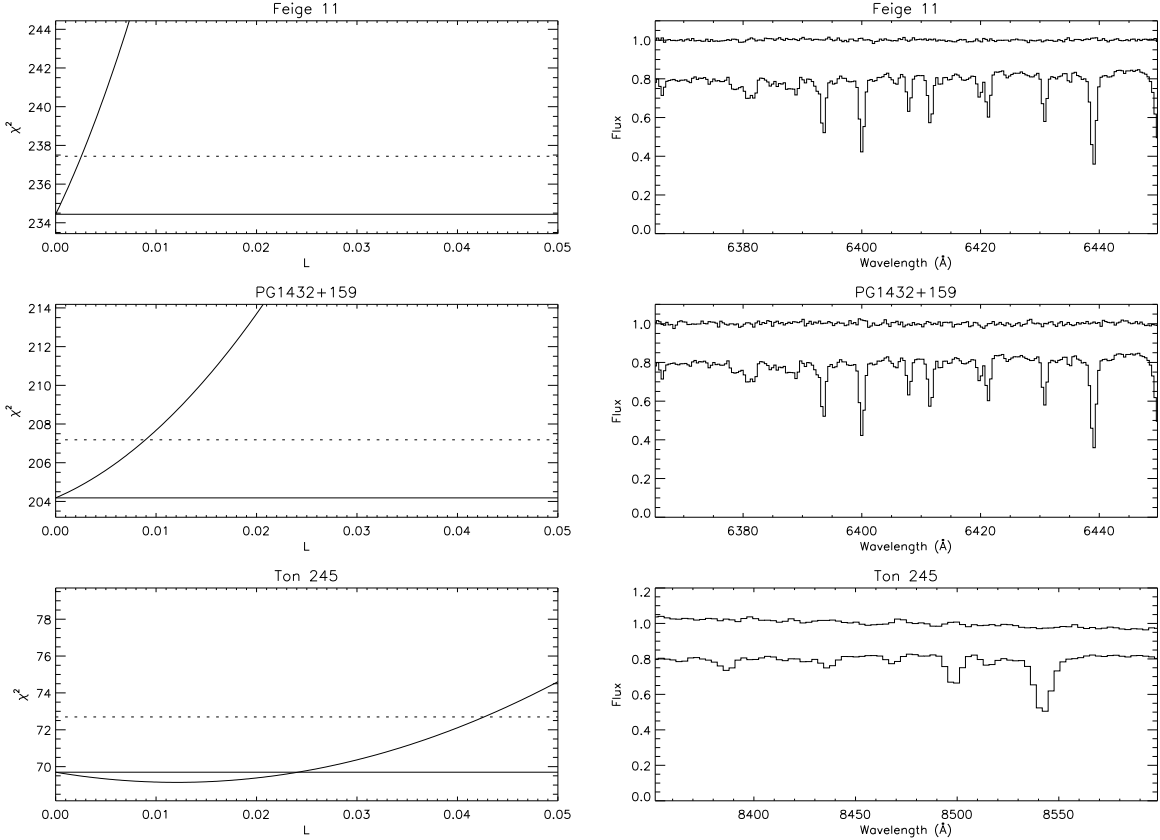,width=0.9\textwidth}
\end{figure*}

 For Feige~11 and  PG\,1432+159 we used the average spectra near H$\alpha$
described in Moran et~al. (1999). We compared these spectra to a spectrum of
GL\,69, a K5V star. All the spectra were shifted in wavelength so that
spectral features appear at their rest wavelengths and then rebinned onto
uniform wavelength grid of 215 elements between 6365\AA\ and 6450\AA. There
are no significant spectral features from the sdB star or the Earth's
atmosphere in this wavelength region. The spectra were normalized to give a
continuum value of 1. We then calculated the $\chi^2$ statistic for a
least-squares parabolic fit to the residual $f_{sdB} - L\cdot f_{GL69} + L$
where $f_{sdB}$ and $f_{GL69}$ are the normalised spectra of the sdB star and
GL\,69 respectively. The results are shown as a function of the luminosity
ratio, $L$, in Fig.~\ref{LratioFig}.

 If $\chi_0^2$ is the value of  $\chi^2$ for $L=0$, then we can set an upper
limit to $L$ by considering the minimum value of $L$ for which $\chi^2$ is
significantly worse than $\chi_0^2$. We have chosen a conservative value of 
$\chi_0^2 + 3$ (91.6\,percent confidence limit) and find corresponding upper
limits of $L <0.0026$ for Feige~11 and $L<0.009$ for PG\,1432+159.

 We applied the same technique to our low-resolution ISIS spectrum of Ton~245
and the K3V GL\,250\,A in the spectral region 8350\AA\ to 8600\AA, which is
also free of spectral features from the sdB stars or the Earth's atmosphere.
The rebinned spectra have 86 pixels in this case. The results are shown in
Fig.~\ref{LratioFig} where it can be seen that $L <0.043$. The limit on $L$ is
less stringent than those calculated for Feige~11 or PG\,1432+158 because
there are fewer spectral features visible at lower resolution. We also applied
the technique to Ton~245 and the M1.5V star GL\,220 and found $L <0.046$.
Although the value of $\chi^2$ is slightly improved by subtracting 1--2
percent of a cool star spectrum from the spectrum of Ton~245, this is not a
significant improvement and it is likely to be due to the cancellation of
flat-fielding errors, sky subtraction problems and weak absorption features
due to the Earth's atmosphere rather than any real detection of a cool
companion.

\subsection{The amplitude of the reflection effect from a main-sequence
companion}
\begin{figure*} 
\caption{\label{AmpFig} The amplitude of the reflection effect, $\delta m$,
predicted by our model  in V (solid line) and I (dashed line) and the
luminosity ratio in the V band, $L_{\rm V}$, as a function of the inclination
for a main-sequence companion. }
\psfig{file=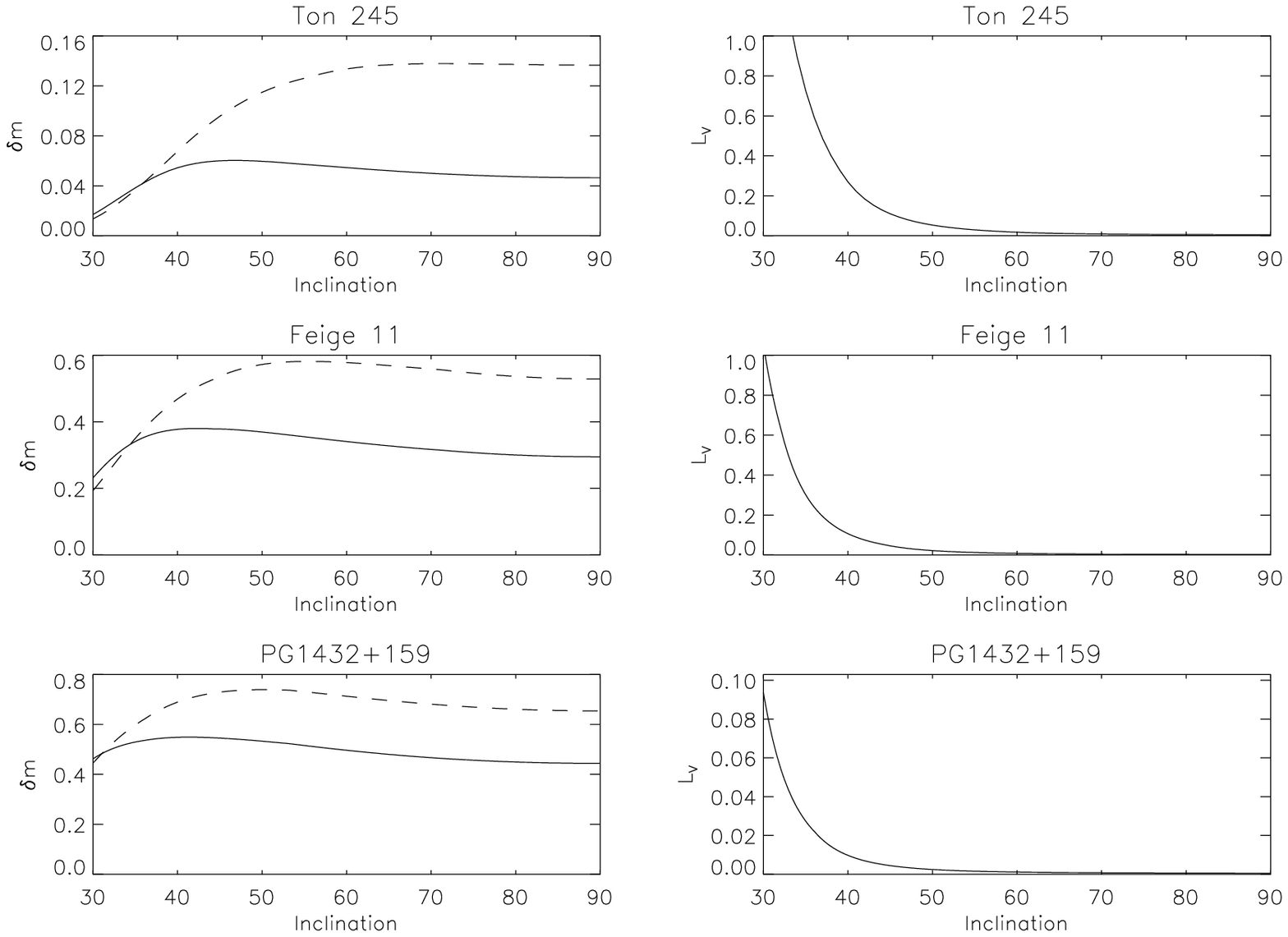,width=0.9\textwidth} 
\end{figure*}

 In order to calculate the  amplitude of the reflection effect from a
main-sequence and the luminosity ratio as a function of inclination, we
proceed as follows.
 For a given  orbital inclination, the semi-amplitudes and orbital periods
given in Table~\ref{DataTable} can be used to compute the mass of the
companion star and the separation of the stars assuming a mass of 0.5\Msolar\
for the sdB star.   The radius of the sdB star can be estimated using the
surface gravity given by Saffer et~al. (1994). The radius, effective
temperature and absolute visual magnitude of the companion star can be
estimated from its mass using the tabulations of Zombeck (1990). We can then
use our model for the reflection effect to predict the amplitude of the
reflection effect as a function of the inclination. The luminosity ratio in
the V-band can also be calculated given the absolute visual magnitude of the
sdB star calculated from the radius and effective temperature of the sdB star
combined with a surface brightness in the V band from the model atmospheres
described above. The results of these calculations are shown in
Fig.~\ref{AmpFig}. We can see from the right-hand panels of Fig.~\ref{AmpFig}
that the upper limit to the luminosity ratio calculated above sets a lower
limit to the inclination of the binary if we assume the companion is a
main-sequence star. Although the luminosity ratio was calculated at redder
wavelengths than the V-band, the cool companion is redder than the sdB star at
these wavelengths, so this is a pessimistic assumption, i.e., we allow a
greater range of inclinations by applying these upper limits calculated from
the spectra to the V-band. In Table~\ref{DataTable} we list the minimum
amplitude of the reflection effect predicted by our model for the range of
inclinations allowed by the upper limit to luminosity ratio, $\delta m_{\rm
min}$. 

\subsection{The observed lightcurves}

\begin{figure*} 
\caption{\label{LCFig} Lightcurves in V and I of Ton~245, Feige~11 and
PG\,1432+159. The points mark the magnitude of the target relative to the
total flux in the companion stars listed in Table~\ref{PhotTable}.
Least-squares fits of a cosine with the same period as the orbital period are
also shown.}
\psfig{file=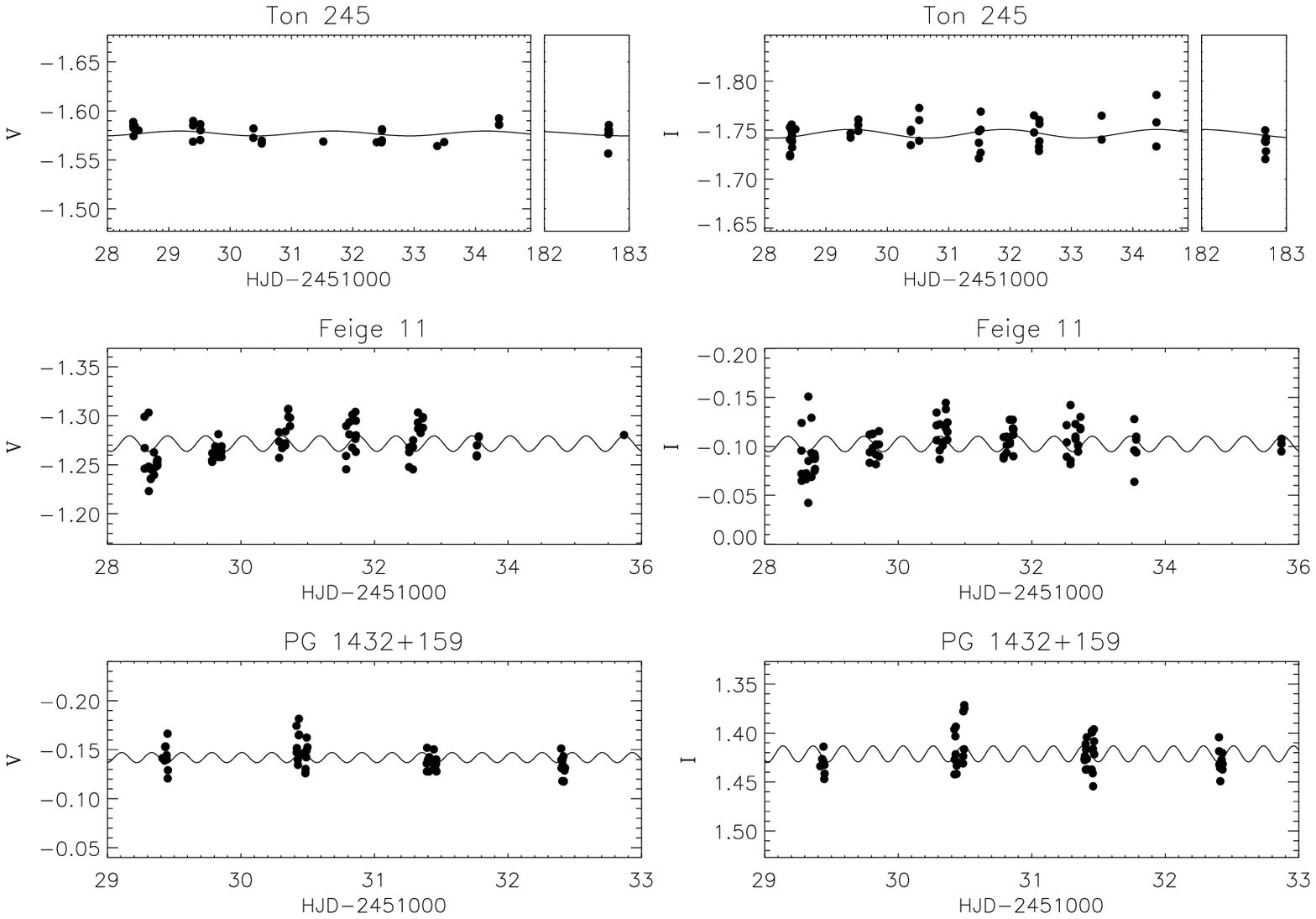,width=0.9\textwidth}
\end{figure*}

\begin{figure*} 
\caption{\label{PerFig} Periodograms of our V and I lightcurves for Ton~245,
Feige~11 and PG\,1432+159. The orbital frequency is marked with a dashed
vertical line. The short horizontal line marks the minimum semi-amplitude 
expected for a non-degenerate companion, $\delta m _{\rm min}/2$.}
\psfig{file=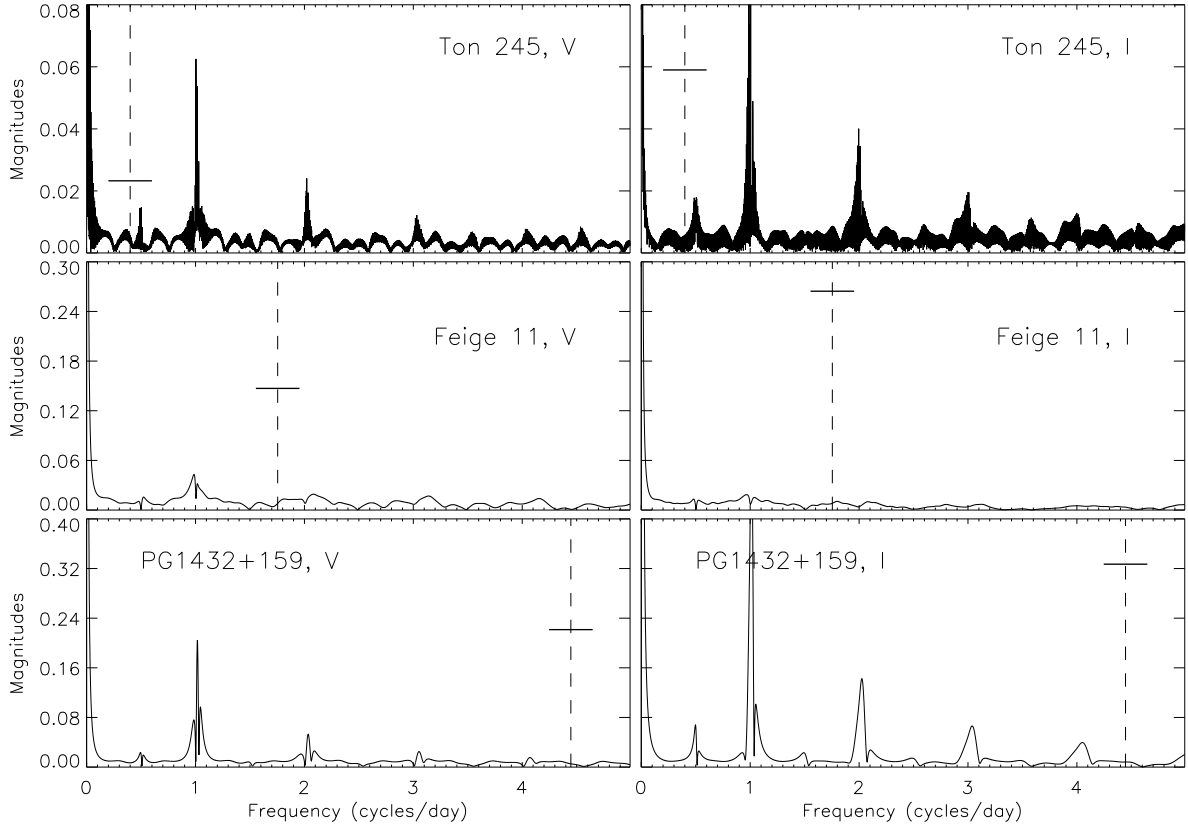,width=0.9\textwidth} 
\end{figure*}

 The lightcurves of Ton~245, Feige~11 and PG\,1432+159 are shown in
Fig.~\ref{LCFig}. Also shown are the results of a least-squares fit of a
cosine wave to these data with the same period as the orbital period. The
amplitude of these cosine waves, $\delta m$, is given in
Table~\ref{LCFitTable} together with the standard deviation of the residuals,
$\sigma_{\delta m}$. We have also calculated periodograms for each lightcurve,
i.e,. the semi-amplitude of sine wave fit by least squares as a function of
frequency. The results are shown in Fig.~\ref{PerFig}. Almost all the power in
these periodograms occurs near 1 cycle/day or its aliases and is due to a
combination of the window function and differential extinction between the
target and comparison stars. It is clear that there is no significant
variability in these lightcurves with the same period as the orbital period
and that any reflection effect in these binaries has an amplitude less than
about 0.01 magnitudes.  We compare these measured semi-amplitudes to the
minimum semi-amplitudes for the reflection effect from a main-sequence
calculated above by calculating the ratio $(\delta m_{\rm min} - \delta
m)/\epsilon_{\delta m}$, where $\epsilon_{\delta m}$ is the uncertainty in
$\delta m$. We see that in all three cases the minimum predicted
semi-amplitude exceeds the observed semi-amplitude by an order-of-magnitude
more than its uncertainty.

 We have also considered the sources of uncertainty in our analysis. The
uncertainties in the properties of the sdB star, i.e., mass, radius and
temperature, affect the value of $\delta _{\rm min}$ by no more than a few
hundredths of a magnitude. Another source of uncertainty is the scatter in
the mass-radius relation for M-dwarfs, which is about 12\% (Caillault \&
Patterson 1990), but this also has a small effect on $\delta _{\rm min}$. The
largest source of uncertainty in our analysis is the error introduced by the
assumptions and approximations used in our model of the reflection effect.
These are difficult to quantify but we have shown that for sdB binaries
similar to those we are studying but with M-dwarf companions, the model is
able to predict the amplitude of the reflection effect to within 50\,percent.
Even if we are pessimistic and assume that the uncertainties in the models
used to derive $\delta m_{\rm min}$ are a factor of a few, it is clear that
the amplitude of any reflection effect in the lightcurves of Ton~245 are a
factor of at least $\sim 4$ lower than that expected from  a main-sequence
companion and are an order-of-magnitude lower for the lightcurves of Feige~11
and PG\,1432+159.

 There is no evidence in any of our lightcurves of variability on timescales
of 90s\,--\,600s, which is the typical period range for pulsating sdB stars
(Koen et~al. 1999). However, our data are far from ideal for studying
pulsations given the long exposure times used and poor data sampling so it is
quite possible that one or more of the stars studied are pulsating sdB stars.

\section{Discussion}
 We can rule out the possibility that the companions to Ton~245, Feige~11 and
PG\,1432+159  are main-sequence stars because they do not show any reflection
effect in their lightcurves. We can also rule out a sub-giant companion in all
three cases because a sub-giant is, by definition, larger than a main-sequence
star, so the reflection effect would be much larger. We conclude that none of
these sdB stars has a main-sequence or sub-giant companion. The companions to
these sdB stars have masses $\ga 0.3\Msolar$ but must have radii much smaller
than main-sequence stars to avoid the reflection effect being seen in the
lightcurve. A white dwarf companion satisfies these constraints comfortably
and sdB\,--\,white dwarf binaries are known to exist, e.g., KPD\,0422+5421
(Koen, Orosz \& Wade 1998), KPD\,1930+2752 (Maxted et~al. 2000). 

 The orbital period we have measured for PG\,1017$-$086  is the shortest period
for an sdB binary measured to-date and is comparable to those of short-period
cataclysmic variables. The timescale for orbital shrinkage due to the loss of
gravitational waves is about 800\,Myr, which is comparable to the lifetime of
an sdB star. It is interesting to speculate what the binary will look like if
mass transfer begins before the sdB star evolves into a low mass white dwarf.
The stream of material from the M-dwarf will impact directly onto the surface
of the sdB star, i.e.,  no accretion disk will form, so the effects of the
mass transfer may not be immediately obvious.

\begin{table}
\caption{\label{LCFitTable} Cosine fits to our V and I lightcurves. The
amplitude of the cosine, $\delta m$, and the standard deviation of the
residuals, $\sigma_{\delta m}$, are given in magnitudes with the number of
observations given in parentheses. The figure in the final column is the ratio
of ($\delta m_{\rm min} - \delta m)/\epsilon_{\delta m}$, where
$\epsilon_{\delta m}$ is the uncertainty in $\delta m$.}
\begin{tabular}{@{}lcrrr}
Name & \multicolumn{1}{l}{Filter} &
\multicolumn{1}{l}{$\delta m$} &
\multicolumn{1}{l}{$\sigma{\delta m}$} &
\multicolumn{1}{l}{$\frac{(\delta m_{\rm min}-\delta m)}{\epsilon_{\delta
m}}$} \\
Ton\,245    & V &       0.0050 & 0.0081 & 10.9 \\
            &   &  $\pm$0.0038 & (n=38)   \\
            & I &       0.0088 & 0.0137 & 19.5 \\
            &   &  $\pm$0.0056 & (n=49)    \\
Feige\,11   & V &        0.0162 & 0.0187 &  46.4\\
            &   &   $\pm$0.0060 & (n=80)   \\
            & I &        0.0162 & 0.0192 &  88.3\\
            &   &   $\pm$0.0058 & (n=85)   \\
PG\,1432+159& V &        0.0104 & 0.0127 & 90.3  \\
            &   &   $\pm$0.0048 & (n=57)   \\
            & I &        0.0164 & 0.0175 & 91.1 \\
            &   &   $\pm$0.0070 & (n=52)  \\
\end{tabular}
\end{table}

\begin{table*}
\caption{\label{DataTable} Parameters of Ton 245, Feige~11 and PG\,1432+159.
The effective temperature (T$_{\rm eff}$) and surface gravity ($\log g$) are
taken from Saffer et~al. (1994).  M$_{\rm min}$ is the minimum mass of the
companion assuming a mass of 0.5\Msolar\ for the sdB star and $\delta m _{\rm
min}$ is the minimum amplitude of the reflection effect expected 
from a main-sequence companion given the upper limit to the luminosity ratio
$L_{\rm max}$.}
\begin{tabular}{@{}lrrrrrrrrl}
Name & T$_{\rm eff}$ & $\log g$ &  \multicolumn{1}{l}{P} &
\multicolumn{1}{l}{K} & \multicolumn{1}{l}{M$_{\rm min}$} & 
\multicolumn{1}{l}{$L_{\rm max}$} &
\multicolumn{2}{c}{$\delta m _{\rm min}$} &Ref. \\
     & (K) & (cgs) &  \multicolumn{1}{l}{($d$)} &
\multicolumn{1}{l}{(\kms)} & \multicolumn{1}{l}{(\Msolar)} && V & I & \\
\hline
Ton 245 &25200 &5.30& 2.5 &75 &0.47 &
0.0046  & 0.05 & 0.12&  Foss, Wade \& Green, 1991\\
Feige~11  & 28400&5.63&  0.569908& 104.3 & 0.37 &0.0026 & 0.29 
& 0.53 & Moran et~al., 1999 \\
PG\,1432+159  &26900 &5.75& 0.2249 & 120.0 & 0.30 &0.009  & 0.44 & 
0.65 & Moran et~al., 1999 \\
\hline
\end{tabular}
\end{table*}

\section{Conclusion}
 
 We have presented lightcurves in two colours of three binary subdwarf B
stars --  PG\,1432+159, Feige~11 and Ton~245. We have shown that there is no
sign in these lightcurves of the sinusoidal variation which would be seen if
the companion star were a main-sequence star or a sub-giant. The most likely
explanation is that all three sdB stars have white dwarf companion stars. 

 By contrast, the reflection effect in PG1017$-$086 is clearly seen in the
lightcurve. We have presented spectroscopy of this star from which we have
measured the orbital period, mass function and projected rotational velocity.
These observations show that PG1017$-$086 is an sdB star with a low-mass
M-dwarf or brown dwarf companion. 

\section*{Acknowledgments}
 PFLM was supported by a PPARC post-doctoral grant. The Jacobus Kapteyn
Telescope is operated on the island of La Palma by the Isaac Newton Group in
the Spanish Observatorio del Roque de los Muchachos of the Instituto de
Astrofisica de Canarias. The SAAO is a National Facility administered by the
National Research Foundation of South Africa.

\label{lastpage}


\begin{thebibliography}{99}
%
\bibitem{Cail90} Caillault J.-P., Patterson J., 1990, AJ, 100, 825
%
\bibitem{Dcru96} d'Cruz N.L., Dorman B., Rood R.T., O'Connell R. W., 1996,
ApJ, 466, 359.
%
\bibitem{Down86} Downes R.A.,  1986, ApJS, 61, 569.
%
\bibitem{Dres01} Drechsel H., Heber U., Napiwotzki R., \O stensen R., 
Solheim J.-E., Johannessen F., Schuh S.L., Deetjen J., Zola S.,  2001, A\&A,
 379, 893.
%
\bibitem{Foss91} Foss D., Wade R.A., Green R.F., 1991, ApJ, 374, 281.
%
\bibitem{Gree86} Green R.F., Schmidt M., Liebert J., 1986, ApJS, 61, 305.
%
\bibitem{Hebe86} Heber U., 1986, A\&A 155, 33.
%
\bibitem{Hild96} Hilditch R.W., Harries T.J., Hill G.,  1996, MNRAS, 279, 1380.
%
\bibitem{Hill93} Hill G., Rucinski, S.M., in Milone E.C., ed., Lightcurve
Modeling of Eclipsing Binary stars. Springer-Verlag, Berlin, p.135.
%
\bibitem{Iben93} Iben I., Livio M., 1993, PASP, 105, 1373.
%
\bibitem{Iria57} Iriarte B., 1959, Lowell Obs. Bull., 4, 130.
%
%
\bibitem{Kilk97}  Kilkenny D., O'Donoghue D., Koen C., Stobie R.S., Chen A.,
1997, MNRAS, 287, 867
%
\bibitem{Kilk98} Kilkenny D., O'Donoghue D., Koen C., Lynas-Gray A.E., 
van~Wyk F., 1998, MNRAS 296, 329.
%
\bibitem{Kiss00} Kiss L.L., Cs\'{a}k B., Szatm\'{a}ry,K., Fur\'{e}sz G.,
Szil\'{a}di K., 2000, A\&A, 364, 199.
%
\bibitem{Koen98} Koen C., Orosz J.A., Wade R.A., 1998, MNRAS, 300, 695.
%
\bibitem{Koen99} Koen C., O'Donoghue D., Pollacco D.L., Charpinet S., 1999, 
MNRAS, 305, 28.
%
\bibitem{Land83} Landolt, A.U., 1983, AJ, 88, 439.
%
\bibitem{Oke99} Oke J.B., 1990, AJ, 99, 1621 
%
\bibitem{Mars89} Marsh, T.R., 1989, PASP 101, 1032.
%
\bibitem{Maxt01} Maxted P.F.L., Heber, U., Marsh T.R., North R.C., 2001,
MNRAS, 326, 1391.
%
\bibitem{Maxt00} Maxted P.F.L., Marsh T.R., North R.C., 2000, MNRAS, 317,
L41.
%
\bibitem{Mora99} Moran C., Maxted P., Marsh T.R., Saffer R.A., Livio, M. 1999,
MNRAS, 304, 535.
%
\bibitem{Napi97} Napiwotzki R., 1997, A\&A, 322, 256
%
\bibitem{Nayl98} Naylor, T., 1998, MNRAS, 296, 339
%
\bibitem{Poll94}Pollacco D.L., Bell S.A.,  1994, MNRAS, 267, 452.
%
%
\bibitem{Saff94} Saffer R.A., Bergeron P., Koester D., Liebert J., 1994, ApJ, 
432, 351.
%
%
\bibitem{Wese92} Wesemael F., Fontaine G., Bergeron P., Lamontagne R, Green
R.F.
 1992, AJ 104, 203
%
\bibitem{Wils71} Wilson R.E., Devinney E.J.,  1971, ApJ, 166, 605.
%
\bibitem{Wood93} Wood J.H., Zhang E.H., Robinson E.L., 1993, MNRAS, 261, 103.
%
\bibitem{Wood99} Wood J.H., Saffer R.,  1999, MNRAS, 305, 820.
%
\bibitem{Zombeck} Zombeck M.V., 1990, Handbook of Space Astronomy and
Astrophysics, Second Edition, Cambridge University Press, Cambridge, UK. 
%
\end{thebibliography}
\end{document}